\documentclass[referee]{raa}           
\usepackage{graphicx,times}
\usepackage{natbib}
\usepackage{amssymb,amsmath}
\bibpunct{(}{)}{;}{a}{}{,}

\usepackage[a4paper=true,dvipdfm=true,pagebackref=true]{hyperref}
\hypersetup{pdftitle = The title of my PDF, pdfauthor = My name, pdfsubject= The subject, pdfkeywords = keyword1 keyword2 keyword3} 
\hypersetup{colorlinks = true, linkcolor = green, anchorcolor = red, citecolor = blue, filecolor = red, pagecolor = red, urlcolor = red}

\begin{document}

   \title{Dynamics of an ensemble of clumps embedded in a magnetized ADAF}

   \setcounter{page}{1}

   \author{Fazeleh Khajenabi\inst{1}, Mina Rahmani \inst{2}, }

   \institute{ Department of Physics, Golestan University, Gorgan 49138-15739, Iran; {\it f.khajenabi@gu.ac.ir }\\
        \and
             Department of Physics, Damghan University, Damghan, Iran\\
\vs \no
}

\abstract{We investigate effects of a global magnetic field on the dynamics of an ensemble of clumps within a magnetized advection-dominated accretion flow by neglecting interactions between the clumps and then solving the collisionless Boltzman equation. In the strong-coupling limit, in which the averaged radial and the rotational velocities of the clumps follow the ADAF dynamics, the averaged radial velocity square of the clumps is calculated analytically for different magnetic field configurations. The value of the averaged radial velocity square of the clumps increases with increasing the strength of the radial or vertical components of the magnetic field. But a purely toroidal magnetic field geometry leads to a reduction of the value of the averaged radial velocity square of the clumps at the inner parts with increasing the strength of this component. Moreover, dynamics of the clumps strongly depends on the amount of the advected energy so that the value of the averaged radial velocity square of the clumps increases in the presence of a global magnetic field  as the flow becomes more advective.}


   \authorrunning{F. Khajenabi \& M. Rahmani}            
   \titlerunning{Dynamics of clumps in ADAFs}  
   \maketitle

%
\section{Introduction}           
\label{sect:intro}

Accretion processes have been extensively studied during last decades and several types of the accretion models have been proposed to explain certain observational features of some of the astrophysical objects. Most of the models assume that the accretion process occurs as one-component gaseous fluid. But there are strong observational and theoretical arguments which imply at least some of accreting systems are clumpy so that they consist of cool clumps embedded in a much hotter and more tenuous gaseous fluid. For example, observational evidences show that broad-line region of active galactic nuclei (AGN) has a clumpy structure (\citealt{Rees+1987}; \citealt{Krolik+Begelman+1988}; \citealt{Nenkova+etal+2002}; \citealt{Risaliti+etal+2011}; \citealt{Torricelli-Ciamponi+eta+2014l} ). The broad emission lines in the spectrum of AGNs are attributed to an assembly of clouds which are moving through a hot intercloud medium. Clouds' basic properties are estimated according to the photoionisation models. These models predict that clouds' typical size is $10^{12\pm  1} {\rm cm}$ and their number density is $10^{10\pm 1} {\rm cm}^{-3}$ (e.g., \citealt{Rees+1987}; \citealt{Krause+Schartmann+Burkert+2012}). Orbital motion of BLR clouds is a rich source of information for estimating the mass of the central black hole (e.g.,\citealt{Netzer+Marziani+1994}). One can neglect collisions between the clumps and investigate the orbit of an individual clump in the presence of the central gravitational force and possible radiation field like a two-body classical problem (e.g., \citealt{Netzer+Marziani+1994}; \citealt{Krause+Burkert+Schartmann+2011}; \citealt{Krause+Schartmann+Burkert+2012}; \citealt{Plewa+Schartmann+Burkert+2013}; \citealt{Khajenab+2015}  ).  Although there are theoretical concerns about the stability of the clumps, it is generally believed  that magnetic fields provide a confinement mechanism (e.g., \citealt{Rees+1987}). 

Another approach to study dynamics of the clumps embedded in a hot medium is based on analyzing the collisionless Boltzmann equation as has been done by \citealt{Wang+Cheng+Li+2012} (hereafter WCL).  They described the gaseous ambient medium using the classical similarity solutions of Advection-Dominated Accretion Flows (ADAFs) presented by \cite{Narayan+Yi+1994} for the non-magnetized systems. Although collisions between the clumps have been neglected for simplicity, their interactions with the surrounding gaseous medium was included through a drag force as a function of the relative velocity of the clumps and the gas. In the strong-coupling limit, it was shown that the root of averaged radial velocity square
of the clumps is much larger than radial velocity of the gas flow. The analysis has been extended to the magnetized case by \cite{Khajenabi+Rahmani+Abbassi+2014} where the authors considered a purely toroidal magnetic field geometry for the gaseous component. They found that when magnetic pressure is less than the gas pressure, the averaged radial velocity of clumps decreases at the inner regions of the system whereas it increases at the outer parts, though this enhancement is not very significant unless the system becomes magnetically dominated.

In this work, we extend the analysis by \cite{Khajenabi+Rahmani+Abbassi+2014}  to include all {\it three components} of the magnetic field in the gaseous component. Since properties of the gas flow is significantly modified in the presence of a global magnetic field \citep{Zhang}, the drag force varies depending on the strength of the magnetic field and the assumed magnetic geometry which will eventually lead to a considerable modification in the velocity dispersion of clumps. In the next section, we present our basic assumptions and the equations. In section 3, a parameter study for the dynamics of the clumps are presented. We conclude with a summary of the results in section 4.

\section{ GENERAL FORMULATION}
Our analysis for describing an ensemble of clumps is based on WCL approach which implements  the collisionless
Boltzman equation in the cylindrical coordinates $ (r, \phi, z) $  including the components of the drag force. If we assume the distribution function of clumps is represented by ${\cal F}$, then Boltzman equation is written as

\begin{equation}
\frac{\partial {\cal F}}{\partial t} + \dot{R} \frac{\partial {\cal F}}{\partial r} + \dot{\phi}  \frac{\partial {\cal F}}{\partial \phi} + \dot{z} \frac{\partial {\cal F}}{\partial z} + \dot{v}_{r} \frac{\partial {\cal F}}{\partial v_r } + \dot{v}_{\phi} \frac{\partial {\cal F}}{\partial v_\phi } + \dot{v}_{z} \frac{\partial {\cal F}}{\partial v_z } + {\cal F} (\frac{\dot{v}_r}{\partial v_r} + \frac{\dot{v}_\phi}{\partial v_\phi} + \frac{\dot{v}_z}{\partial v_z}) =0,
\end{equation}
where $\dot{r} = v_r$, $\dot{\phi}=v_\phi / r$ and $\dot{z}=v_z$. The central object with mass $M$ is at the origin and the gravitational potential becomes $ \Phi=GM/(r^{2}+z^{2})^{1/2} $. The components of the drag force are $ F_{r}=f_{r}(v_{r}-V_{r})^{2} $ and $ F_{\phi}=f_{\phi}(v_{\phi}-V_{\phi})^{2} $ where $ f_{r} $ and $ f_{\phi} $ are constants of order unity. Here, the radial and the rotational velocities of the ADAF where clumps are moving within it are denoted by $ V_{r} $ and $ V_{\phi} $. Thus, dynamical properties of the background medium affect dynamics of the clumps through these components of the velocity. Thus, Boltzman equation becomes (WCL)
\begin{equation}
\frac{\partial {\cal F}}{\partial t} + v_r \frac{\partial {\cal F}}{\partial r} + v_z \frac{\partial {\cal F}}{\partial z} + \left( \frac{v_{\phi}^2}{r} - \frac{\partial \Phi}{\partial r} + F_r \right) \frac{\partial {\cal F}}{\partial v_r} + \left( F_\phi - \frac{v_r v_\phi }{r} \right) \frac{\partial {\cal F}}{\partial v_\phi} - \frac{\partial\Phi}{\partial z} \frac{\partial {\cal F}}{\partial v_z} + 2 {\cal F} \left[ F_\phi (v_\phi - V_\phi ) + F_r (v_r - V_r) \right]=0.
\end{equation}
It is very unlikely to solve this equation analytically in a general case unless we apply further simplifying assumptions. We assume that  the clumps are strongly coupled with the background gaseous medium which implies the mean radial and the rotational velocities of the clumps are equal to the radial and the rotational velocity of ADAF. Under these simplifying assumptions, it is possible to obtain the root mean radial velocity square of the clumps $ \langle v_{r}^{2}\rangle^{1/2} $ analytically (WCL): 
   \begin{equation}
\langle v_{r}^{2} \rangle=c^{2}\lbrace \frac{1}{2}[\alpha^{2}c_{1}^{2}\Gamma_{r}\Lambda_{\frac{3}{2}}(\Gamma_{r},r)+(1-c_{2}^{2})\Lambda_{\frac{5}{2}}(\Gamma_{r},r) -\frac{\alpha c_{1}c_{2}}{2\Gamma_{\phi}}\Lambda_{\frac{7}{2}}(\Gamma_{r},r)]+\frac{V_{out}^{2}}{c^{2}r_{out}^{\frac{1}{2}}}e^{-\Gamma_{r}r_{out}}\rbrace\times r^{\frac{1}{2}}e^{\Gamma_{r}r},
\label{eq:vr}
\end{equation}
where $ \Gamma_{r}=f_{r}R_{sch} $ and $ \Gamma_{\phi}=f_{\phi}R_{sch} $ are the coefficients of the drag force. Function $ \Lambda_{q} $ is introduced by WCL as $ \Lambda_{q}=\int^{r_{out}}_{r} x^{-q}exp(-\Gamma_{R}x)dx $.  The outer boundary condition is at $ r=r_{out} $, so that  $ \langle v_{r}^{2}\rangle=V_{out}^{2} $.
Moreover, properties of the ADAF are described using a set of radially self-similar solutions where $ c_{1} $ and $ c_{2} $ are coefficients of the radial and the rotational velocities of the ADAF. In WCL, the standard nonmagnetized ADAF solutions \citep{ Narayan+Yi+1994} have been used for their analysis. Then, \cite{Khajenabi+Rahmani+Abbassi+2014}  extended the analysis by including purely toroidal component of the magnetic field using similarity solutions of \cite{Akizuki+Fukue+2006}. But we extend previous studies by considering all three components of the magnetic field using similarity solutions of \cite{ Zhang}. Magnetized self-similar solutions of \cite{Zhang} are written as

\begin{equation}
v_{r}(r)=-c_{1}\alpha \sqrt{\frac{GM}{r}},
\end{equation}

\begin{equation}
v_{\phi}(r)=c_{2}\sqrt{\frac{GM}{r}},
\end{equation}

\begin{equation}
c_{s}^{2}(r)=c_{3}\frac{GM}{r},
\end{equation}

\begin{equation}
c^{2}_{r,\phi ,z}(r)=\frac{B^{2}_{r,\phi ,z}}{4\pi \rho}=2\beta_{r,\phi ,z}c_{3}\frac{GM}{r},
\end{equation}
where the coefficients $ \beta_{r} $, $ \beta_{\phi} $, $ \beta_{z} $ measure the ratio of the magnetic pressure in three directions to the gas pressure, i.e. $ \beta_{r,\phi,z}=P_{mag,r,\phi,z}/P_{gas} $. The coefficients $ c_{1} $, $ c_{2} $ and $ c_{3} $ are obtained using a set of algebraic equations (\citealt{ Zhang}):
\begin{equation}
-\frac{1}{2} c^{2}_{1} \alpha^{2}=c^{2}_{2}-1-[(s-1)+\beta_{z}(s-1)+\beta_{\phi}(s+1)]c_{3}, 
\end{equation}
\begin{equation}
-\frac{1}{2}c_{1}c_{2}\alpha =-\frac{3}{2}\alpha (s+1)c_{2}c_{3}+c_{3}(s+1)\sqrt{\beta_{r}\beta_{\phi}},
\end{equation}

\begin{equation}
c_{2}^{2}=\frac{4}{9f}(\frac{1}{\gamma -1}+s-1)c_{1},
\end{equation}
Here, $ \gamma $ and $ s $ are adiabatic index of the gas and mass loss parameter, respectively. Also, $f$ measures the degree to which the flow is advection dominated. Now, we can substitute the above magnetized self-similar solutions into the equation (\ref{eq:vr}) to study dynamics of the clumps in the presence of a global magnetic field.

\section{ANALYSIS}
We now study the root mean radial velocity square of the clumps $ \langle v_{r}^{2}\rangle^{1/2} $ as a function of the radial distance for different values of the input parameters using the main equation (\ref{eq:vr}). In all Figures, we assume coefficient of the viscosity is $ \alpha =0.1 $  and the adiabatic index is $ \gamma =1.4$.  Moreover, the mass of the central object is fixed at one solar mass. The coefficients of the drag force are $ \Gamma_{r}=5\times 10^{-2} $ and $ \Gamma_{\phi}=2.8\times 10^{-3} $.

Figure 1 shows the root mean radial velocity square of the clumps as a function of the radial distance normalized by $ R_{sch} $ for different values of $ \beta_{r} $ whereas the other magnetic parameters are fixed as $ \beta_{z}=\beta_{\phi}=1 $.  This Figure indicates that the  value of $ \langle v_{r}^{2} \rangle^{1/2} $ increases with $ \beta_{r} $,  although its variation is not very significant.

In Figure $ 2 $, we assume that the radial component of the magnetic field does not exist and the toroidal component is fixed, i.e. $ \beta_{r}=0 $ and $ \beta_{\phi}=1 $.  We can then vary the parameter $ \beta_{z} $ to study its effect on the radial dynamics of the clumps. Again, we see that the value of $ \langle v_{r}^{2}\rangle^{1/2} $ increases as the vertical component of the magnetic field becomes stronger, though its variation with $ \beta_{z} $ is less significant at large values of $ \beta_{z} $. Moreover, at the inner parts of the system clumps are radially moving faster as the parameter $ \beta_{z} $ increases.

Dependence of the root mean radial velocity square of the clumps on the variations of the the toroidal component of the magnetic field is more complicated as it has already been explored by (\citealt{Khajenabi+Rahmani+Abbassi+2014}) for a purely toroidal configuration.  In Figure $ 3 $, we assume that $ \beta_{r}=0 $ and $ \beta_{z}=1 $,  but different values of  $ \beta_{\phi} $ are considered. In comparison to the previous study \citep{Khajenabi+Rahmani+Abbassi+2014},  here, the vertical component of the magnetic field is considered too. The value of the root mean radial velocity square of the clumps decreases at the inner parts of the system with $ \beta_{\phi} $  whereas the value of  $ \langle v_{r}^{2}\rangle^{1/2} $ increases at the outer parts of the system.

Since the radial and the rotational velocities of the gas component strongly depend on the amount of the advected energy, then obviously the clumps experience different values of the drag force depending on the advection parameter $ f $.  We explore dependence of $ \langle v_{r}^{2}\rangle^{1/2} $ on the parameter $ f $ for different magnetic field configurations in Figure $ 4 $. For purely radial or toroidal magnetic field geometries, the value of $ \langle v_{r}^{2}\rangle^{1/2} $ strongly increases with the amount of the advected energy. But for purely vertical magnetic field, this trend changes to a reduction of the mean velocity square of the clumps with increasing the parameter $ \beta_{z} $.  However, this reduction is not very significant. Thus, one may conclude that the value of  $ \langle v_{r}^{2}\rangle^{1/2} $ generally increases as the flow becomes more advective even in the presence of all three components of the magnetic field (Figure 5).

\begin{figure}
   \centering
   \includegraphics[width=10cm, angle=0]{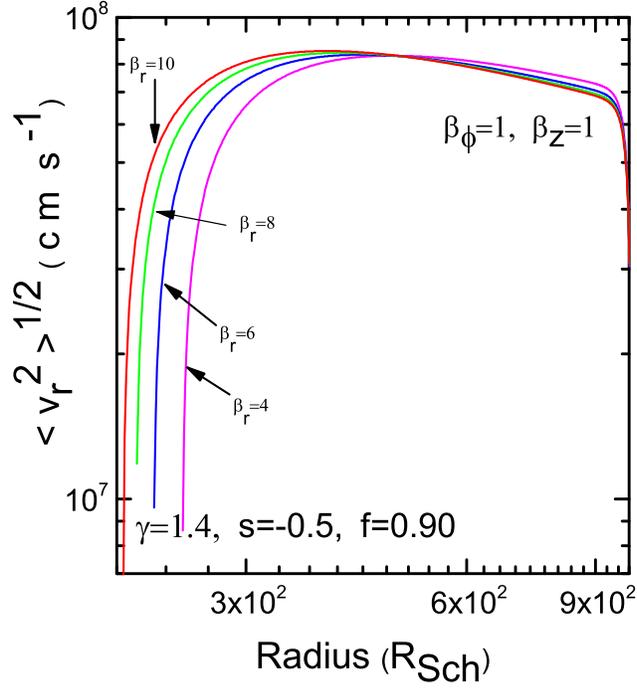}
   \caption{ Root of the averaged radial velocity square  $ \langle v_{r}^{2} \rangle $ of clumps versus the radial distance for a central object with one solar mass. Different values for the parameter $ \beta_{r} $ are considered and each curve is labeled by the corresponding value of this parameter. The rest of the input parameters are $\gamma  = 1.4 $, $s =- 0.5$, $f = 0.9$, $ \beta_{z}=1 $ and $ \beta_{\phi}=1 $.}
   \end{figure}

In all of the figures, most of the curves cannot extend to the very inner parts. It is actually because of the limitation of similarity solutions. We described the gaseous component using self-similar solutions and as we know similarity solutions are valid only at the regions far from the boundaries. In other words, these similarity solutions for gas component are not valid at the very inner parts. But at the intermediate regions, similarity solutions represent dynamics of the gas flow with a very good accuracy. Since dynamics of clumps in our model is determined mainly due to interaction of clumps with the gas and similarity solutions for the gas are not valid at the inner boundary, we not investigate fate of the clumps at the inner parts based on our solutions.
   
\section{conclusions} 
We studied dynamics of an ensemble of cold clumps embedded in a hot magnetized accretion flow. Although magnetic field has a vital role in stability and confinement of cold clouds, their role on the orbital motion of these clumps has not been studied. In our work, properties of the gas component is modified in the presence of a global magnetic field and so, the drag force on each clump changes accordingly. Comparing to the previous study by \cite{Khajenabi+Rahmani+Abbassi+2014}  who assumed the toroidal component of the magnetic field is dominant, we showed that both the radial and the vertical components of the magnetic field also lead to some changes in the averaged radial velocity square of clumps. The value of $ \langle v_{r}^{2}\rangle^{1/2} $ increases with increasing the strength of the radial and the vertical components of the magnetic field. Moreover, velocity dispersion of clumps increases as the flow becomes more advective when all components of the magnetic field are considered. Although results of our analysis are not directly applicable to the real systems because of limitations of this simplified model, the present study clearly demonstrates the importance of the magnetic field in the dynamics of clumps which can not be neglected. The results of the paper are obtained within the conditions of strong
coupling and simplification of the magnetic field. It is also possible to relax these simplifying assumptions but then it would be very unlikely to obtain analytical solutions which is our goal in the present study.

As the clumps move toward the central black hole, they will gradually accumulate at the inner parts because of the tidal disruption of the black hole's gravitational field. In fact, tidal disruption determines the inner edge of the clumpy disc. In the presence of a global magnetic field, we find that on the average the clumps are radially moving faster in comparison to a similar configuration but without magnetic field. WCL calculated the capture rate of clumps and found that it is directly proportional to the ratio of $ \langle v_{r}^{2}\rangle^{1/2} / V_r $. Thus, presence of a global magnetic field increases the capture rate of clumps, but the level of enhancement depends on the detailed input parameters as we explored in this study. In other words, capturing clumps is faster when magnetic fields are considered. 

\begin{figure}
   \centering
   \includegraphics[width=10cm, angle=0]{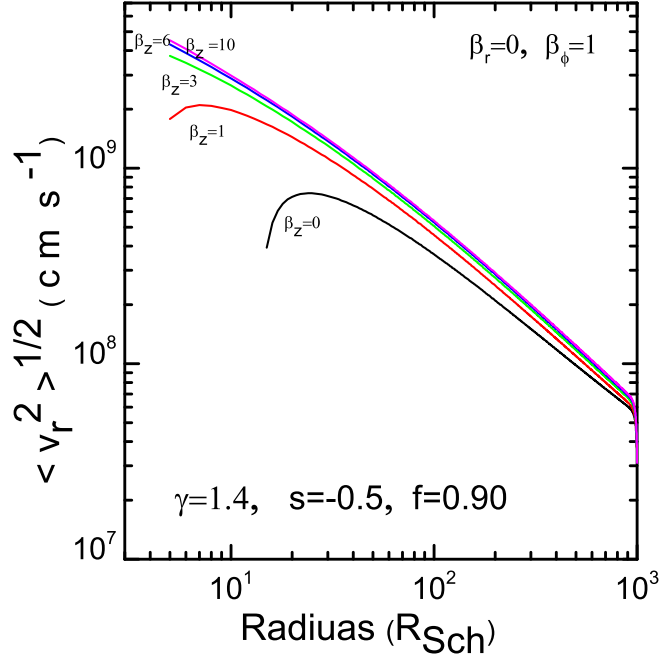}
   \caption{Same as Figure 1, but for different values of $ \beta_{z} $ . Here, we have $ \beta_{r}=0 $ and $ \beta_{\phi}=1 $.  }
   \end{figure}
   
   \begin{figure}
   \centering
   \includegraphics[width=10cm, angle=0]{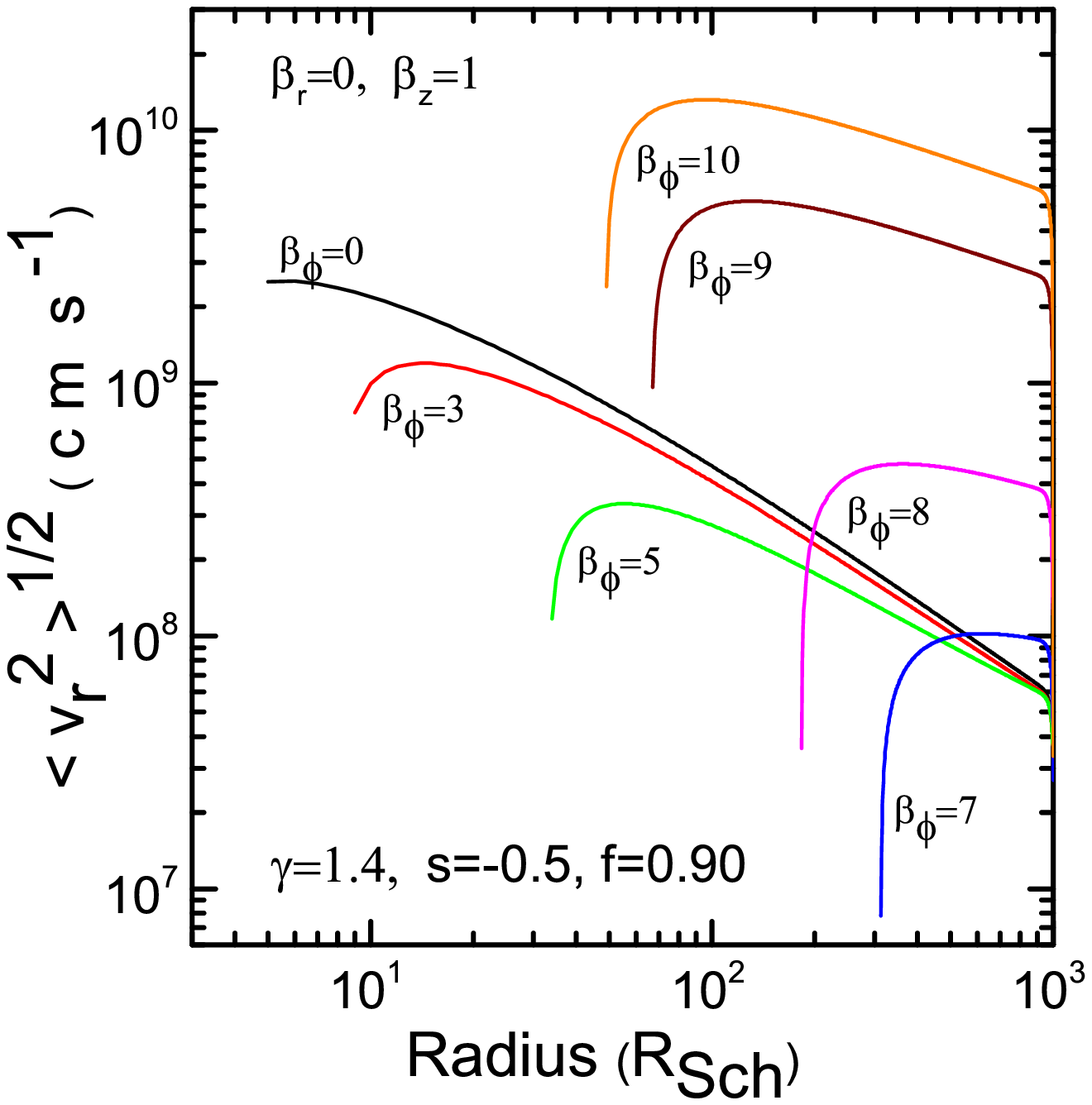}
   \caption{Same as Figure 1, but for different values of $ \beta_{\phi} $ . Here, we have $ \beta_{r}=0 $ and $ \beta_{z}=1 $.  }
   \end{figure}

    \begin{figure}
   \centering
   \includegraphics[width=7cm, angle=0]{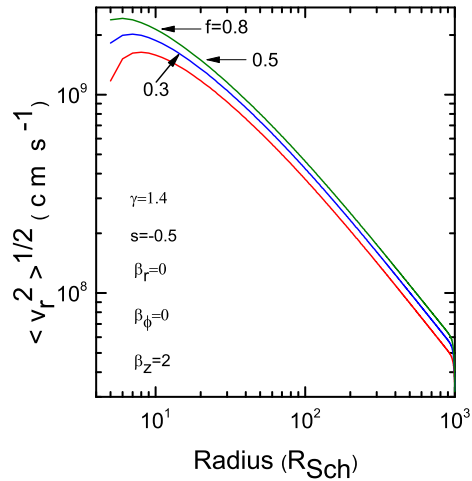}
   \caption{Root of the averaged radial velocity square $ \langle v_{r}^{2}\rangle^{1/2} $ of clumps versus the radial distance for a central object with one solar mass Here, we explore dependence of the root of the averaged radial velocity square on the amount of the advected energy for different magnetic field configurations.}
   \end{figure}
   
   \begin{figure}
   \centering
   \includegraphics[width=10cm, angle=0]{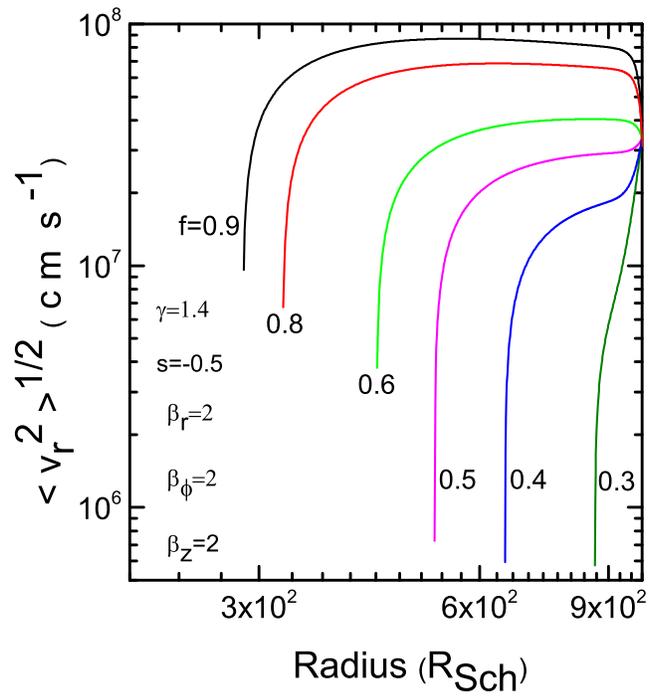}
   \caption{Same as Figure 4,  but all three components of the magnetic field are considered.  }
   \end{figure}

\bibliographystyle{raa}
\bibliography{bibtex}

\begin{thebibliography}{15}
\providecommand{\natexlab}[1]{#1}
\providecommand{\selectlanguage}[1]{\relax}

\bibitem[{{Akizuki} \& {Fukue}(2006)}]{Akizuki+Fukue+2006}
{Akizuki}, C., \& {Fukue}, J. 2006, \pasj, 58, 469

\bibitem[{{Khajenabi}(2015)}]{Khajenab+2015}
{Khajenabi}, F. 2015, \mnras, 446, 1848

\bibitem[{{Khajenabi} et~al.(2014){Khajenabi}, {Rahmani}, \&
  {Abbassi}}]{Khajenabi+Rahmani+Abbassi+2014}
{Khajenabi}, F., {Rahmani}, M., \& {Abbassi}, S. 2014, \mnras, 439, 2468

\bibitem[{{Krause} et~al.(2011){Krause}, { Burkert}, \&
  {Schartmann}}]{Krause+Burkert+Schartmann+2011}
{Krause}, M., { Burkert}, A., \& {Schartmann}, M. 2011, \mnras, 411, 550

\bibitem[{{Krause} et~al.(2012){Krause}, { Schartmann}, \&
  {Burkert}}]{Krause+Schartmann+Burkert+2012}
{Krause}, M., { Schartmann}, M., \& {Burkert}, A. 2012, \mnras, 425, 3172

\bibitem[{{Krolik} \& {Begelman}(1988)}]{Krolik+Begelman+1988}
{Krolik}, J.~H., \& {Begelman}, M.~C. 1988, \apj, 329, 702

\bibitem[{{Narayan} \& {Yi}(1994)}]{Narayan+Yi+1994}
{Narayan}, R., \& {Yi}, I. 1994, \apjl, 428, L13

\bibitem[{{Nenkova} et~al.(2002){Nenkova}, { Ivezic}, \&
  {Elitzur}}]{Nenkova+etal+2002}
{Nenkova}, M., { Ivezic}, Z., \& {Elitzur}, M. 2002, \apjl, 570, L9

\bibitem[{{Netzer} \& {Marziani}(2010)}]{Netzer+Marziani+1994}
{Netzer}, H., \& {Marziani}, P. 2010, \apj, 724, 318

\bibitem[{{Plewa} et~al.(2013){Plewa}, { Schartmann}, \&
  \&~{Burkert}}]{Plewa+Schartmann+Burkert+2013}
{Plewa}, P.~M., { Schartmann}, M., \& \&~{Burkert}, A. 2013, \mnras, 431, L127

\bibitem[{{Rees}(1987)}]{Rees+1987}
{Rees}, M.~J. 1987, \mnras, 228, 47P

\bibitem[{{Risaliti} et~al.(2011){Risaliti}, {Nardini}, {Salvati}
  et~al.}]{Risaliti+etal+2011}
{Risaliti}, G., {Nardini}, E., {Salvati}, M., et~al. 2011, \mnras, 410, 1027

\bibitem[{{Torricelli-Ciamponi} et~al.(2014){Torricelli-Ciamponi}, {Pietrini},
  {Risaliti}, \& {Salvati}}]{Torricelli-Ciamponi+eta+2014l}
{Torricelli-Ciamponi}, G., {Pietrini}, P., {Risaliti}, G., \& {Salvati}, M.
  2014, \mnras, 442, 2116

\bibitem[{{Wang} et~al.(2012){Wang}, {Cheng}, \& \&{ Li}}]{Wang+Cheng+Li+2012}
{Wang}, J.-M., {Cheng}, C., \& \&{ Li}, Y.-R. 2012, \apj, 748, 147, 13

\bibitem[{{Zhang} \& {Dai}(2008)}]{Zhang}
{Zhang}, D., \& {Dai}, Z.~G. 2008, \mnras, 388, 1409

\end{thebibliography}

\end{document}